\documentclass[12pt]{article}

\usepackage{amsmath}
\usepackage{amssymb}
\usepackage[american]{babel}
\usepackage{fourier}
\usepackage[T1]{fontenc}
\usepackage{graphicx}
\usepackage{hyperref}
\usepackage[utf8x]{inputenc}
\usepackage[numbers,sort&compress]{natbib}

\hypersetup{
    colorlinks=true,linkcolor=blue,citecolor=blue,
    filecolor=blue,urlcolor=blue,breaklinks=true
  }

\title{
  The DKP oscillator with a linear interaction in the cosmic string
  space-time
}

\author{
  Mansoureh Hosseinpour$^{1}$%
  \footnote{E-mail:hosseinpour.mansoureh@gmail.com},
  Hassan Hassanabadi$^{1}$%
  \footnote{E-mail:hha1349@gmail.com},\\
  and
  Fabiano M. Andrade$^{2}$%
  \footnote{E-mail:fmandrade@uepg.br}\\
  $^{1}$
  {\small
  Physics Department,
  Shahrood University of Technology,}\\
  \small{Shahrood, Iran,
  P.O. Box 3619995161-316}\\
  $^{2}$
  {\small
  Departamento de Matem\'{a}tica e Estat\'{i}stica,
  Universidade Estadual de Ponta Grossa,}\\
  \small{84030-900 Ponta Grossa, Paran\'{a}, Brazil}
  }

\begin{document}

\maketitle

\begin{abstract}
We study the relativistic quantum dynamics of a DKP oscillator field
subject to a linear interaction in cosmic string space-time in order to
better understand the effects of gravitational fields produced by
topological defects on the scalar field. 
We obtain the solution of DKP oscillator in the cosmic string background.
Also, we solve it with an ansatz in presence of linear interaction.
We obtain the eigenfunctions and the energy levels of the relativistic
field in that background.
\end{abstract}

\begin{small}
  \textbf{Key Words}:
  DKP oscillator, cosmic string, linear potential, ansatz solution.
\end{small}

\begin{small}
\textbf{PACS}: 03.65.Pm, 04.62. +v, 04.20.Jb
\end{small}

\newpage

\section{Introduction}
The Dirac equation including the linear harmonic potential was initially
studied by Ito et al. \cite{NCA.51.1119.1967}, Cook
\cite{LNC.1.419.1971} and Ui et al.\cite{PTEP.72.266.1984}.
This system was latterly called by Moshinsky and Szczepaniak as Dirac
oscillator \cite{JPA.22.817.1989}, because it behaves as an harmonic
oscillator with a strong spin-orbit coupling in the non-relativistic
limit.
This model is based on the dynamics of a harmonic oscillator for
spin-1/2 particles by introducing a nonminimal prescription into free
Dirac equation \cite{JPA.22.817.1989}.
Physically, it can be shown that the Dirac oscillator interaction is a
physical system, which can be interpreted as the interaction of the
anomalous magnetic moment with a linear electric field
\cite{JMP.33.1831.1992,JPA.22.821.1989}.

As a relativistic quantum mechanical system, the Dirac oscillator has
been widely studied. Because it is an exactly solvable model, several
investigations have been developed in the context of this theoretical
framework in the last years. Although the Dirac oscillator is normally
introduced within the context of many body theory, relativistic quantum
mechanics and quantum chromodynamics (in particular as an interquark
potential and also as the confining part of the phenomenological Cornell
potential).
The interest in this issue appears in different contexts
such as quantum optics
\cite{APN.331.120.2013,PRA.77.063815.2008,PRA.76.041801.2007},
supersymmetry \cite{JMP.33.1831.1992,PRL.64.1643.1990,PRD.43.544.1991},
nuclear reactions \cite{PRC.80.044607.2009}, the hadronic spectrum using
the two-body Dirac oscillator\cite{Book.1996.Moshinsky,FP.23.197.1993} a
new representation for its solutions using the Clifford algebra
\cite{PLA.372.2587.2008,JMP.34.4428.1993}, noncommutative
space\cite{JMP.55.32105.2014,IJTP.49.1699.2010}, thermodynamic
properties \cite{PLA.311.93.2003} Lie Algebra symmetries
\cite{JPA.23.2263.1990}, supersymmetric (non-relativistic) quantum
mechanics \cite{PRD.42.1255.1990} the super symmetric path integral
formalism \cite{EPJC.49.1091.2007} the chiral phase transition in
presence of a constant magnetic field \cite{PRA.77.063815.2008},the
relativistic Landau in presence of external magnetic field
\cite{PLA.374.1021.2010} the Aharonov-Bohm effect
\cite{PLA.325.21.2004,EPJC.74.3187.2014} condensed matter physical
phenomena and graphene \cite{ACP.1334.249.2011}. 

The DKP oscillator is an analogous to Dirac oscillator
\cite{JPA.22.817.1989}.
The DKP oscillator in the (1+2)-dimensional noncommutative phase space
for spin-zero particles has been investigated in the work of Guo et
al. \cite{CJP.87.989.2009}. Yang et al. studied the DKP oscillator with
spin-zero in three-dimensional noncommutative phase space
\cite{IJTP.49.644.2010}. 
A generalized bosonic oscillator within the minimal length quantum
mechanics has been analyzed in \cite{JMP.51.33516.2010}.
De Melo et al. released a higher-dimensional formulation of Galilean
covariance to consider the noncommutative DKP oscillator
\cite{JPCS.343.12028.2012}. 
Falek and Merad presented both spin-zero and spin-one DKP equations in
noncommutative space in the (1 + 3)-dimensional case
\cite{CTP.50.587.2008}.
Recently, there has been an increasing interest on the so-called DKP
oscillator
\cite{ZPCPF.56.421.1992,JPA.31.3867.1998,JPA.31.6717.1998,
  JPA.27.4301.1994,PLA.346.261.2005,JMP.47.62301.2006,
  PS.76.669.2007,PS.78.45010.2008,JMP.49.22302.2008,
  IJTP.47.2249.2008}.
The DKP oscillator considering minimal length
\cite{JMP.50.23508.2009,JMP.51.33516.2010}, noncommutative  phase
space
\cite{CTP.50.587.2008,CJP.87.989.2009,IJTP.49.644.2010,EPJC.72.2217.2012}
and topological defects \cite{EPJC.75.287.2015}.

The DKP oscillator embedded in a magnetic cosmic string background has
inspired a great deal of research in last years.
A cosmic string is a linear defect that changes the topology of the
medium when viewed globally.
The spacetime around a cosmic string is locally flat but not globally.
The theory of general relativity predicts that gravitation is manifested
as curvature of spacetime.
This curvature is characterized by the Riemann tensor.
There are connections between topological properties of the space and
local physical laws.
The nontrivial topology of spacetime, as well as its curvature, leads to
a number of interesting gravitational effects.
For example, it has been known that the energy levels of an atom placed
in a gravitational field will be shifted as a result of the interaction
of the atom with spacetime curvature.
Therefore, we have to consider the topology of the spacetime in order to
describe completely the physics of system.

In this work, we examine the relativistic quantum dynamics of DKP
oscillator in the presence of the linear interaction, on the curved
spacetime of a cosmic string. 
From the corresponding DKP equation, we analyze the influence of the
topological defect on the equation of motion, the energy spectrum and
the wave-function. In Sec. 2, we introduce the covariant DKP equation. 
In Sec. 3 we present the covariant DKP oscillator in cosmic string
background and obtain the soloution of DKP oscillator  
In Sec. 4, we present solution of DKP oscillator presence  linear
interaction. In the next section we present our conclusions.


\section{Covariant form of the DKP equation in the cosmic
  string background}

The cosmic string spacetime with an internal magnetic field in
cylindrical coordinates is described by the line element (units such
that $\hbar=c=1$) \cite{PR.121.263.1985,GRG.17.1109.1985}
\begin{align}
  \label{1}
  d{s^2} =  - d{t^2} + d{r^2} + {\alpha ^2}{r^2}d{\varphi ^2} + d{z^2},
\end{align}
with $-\infty < z < \infty$, $\rho \ge 0$ and $0 \le \varphi \le 2\pi$.
The angular parameter $0<\alpha<1$ is related to the linear mass density
$\mu$ of the string as $\alpha  = 1 - 4\mu$.
From the geometrical point of view, the metric  \eqref{1} describes a
Minkowski space{-}time with a conical singularity.

The DKP equation in the cosmic string spacetime \eqref{1}
reads \cite{PRSA.173.91.1939,PR.54.1114.1938,Phdthesis.1936.Petiau}
\begin{align}\label{2}
(i{\beta ^\mu }{\nabla_\mu } - M)\psi  = 0.
\end{align}
The covariant derivative in \eqref{2} is given by \cite{CJP.5.465.2003}  
\begin{align}\label{3}
{\nabla_\mu } = {\partial_\mu } - {\Gamma_\mu }\left( x \right),
\end{align}
where ${\Gamma_\mu }$  are the spinorial affine connections given by
\begin{align}\label{4}
{\Gamma_\mu } = \frac{1}{2}{\omega_{\mu ab}}\left[ {{\beta ^a},{\beta ^b}} \right].
\end{align}
The matrices ${\beta ^a}\left( x \right)$ are the standard Dirac
matrices in Minkowski spacetime  and $\omega_{\mu b}^a$  represents
the spin connection given by 
\begin{align}\label{5}
{\omega_{\mu \bar a\bar b}} = {\eta_{\bar a\,\bar c}}e_\nu ^{\bar c}e_{\bar b}^\sigma \Gamma_{\sigma \mu }^\nu  - {\eta_{\bar a\,\bar c}}e_{\bar b}^\nu e_\nu ^{\bar c}.
\end{align}
The nonnull components of the spin connection are
\begin{align}\label{6}
\omega_\varphi ^{12} =  - \omega_\varphi ^{21} = \,1 - \alpha .
\end{align}

We can build the local reference frame through a non-coordinate basis
with $e_\mu ^{\bar a}$  where $e_\mu ^{\bar a}$ and $e_{\bar a}^\mu $
are transformation matrices.
The components of the non-coordinate basis $e_\mu ^{\bar a}$ are called
tetrads or vierbeins that form our local reference frame.
With the line element \eqref{1}, we can use tetrads $e_{\bar a}^\mu $
and $e_\mu^{\bar a}$ as follows 
\begin{equation}
  \label{7}
  e_{\bar a}^\mu =
  \begin{pmatrix}
    1 & 0 & 0 & 0\\
    0 & \cos \varphi & \sin \varphi & 0\\
    0 & \frac{-\sin\varphi}{\alpha r} & \frac{\cos\varphi}{\alpha r} & 0\\
    0 & 0 & 0 & 1
  \end{pmatrix},
  \qquad
  e_\mu ^{\bar a} =
  \begin{pmatrix}
    1 & 0 & 0 & 0\\
    0 & \cos \varphi & -\alpha r\sin \varphi & 0\\
    0 & \sin \varphi & \alpha r\cos \varphi & 0\\
    0 & 0 & 0 & 1
  \end{pmatrix}.
\end{equation}
The vierbeins form our local reference frame that satisfy the orthonormality conditions
\begin{align}\label{8}
\begin{array}{l}
e_a^\mu \left( x \right)e_\nu ^a\left( x \right) = \delta_\nu ^\mu, \\
e_\mu ^a\left( x \right)e_b^\mu \left( x \right) = \delta_b^a
\end{array}.
\end{align}
And satisfy
\begin{align}\label{9}
{g_{\mu \nu }}\left( x \right) = e_\mu ^{\left( a \right)}\left( x \right)e_\nu ^{\left( b \right)}\left( x \right){\eta_{^{ab}}}.
\end{align}
The Kemmer matrices in curved spacetime are related to their Minkowski
counterparts via 
\begin{align}\label{10}
{\beta ^\mu }\left( x \right) = e_a^\mu \,{\beta ^a}.
\end{align}
In terms of the Minkowski flat spacetime coordinates, these matrices
can be cast into the form 
\begin{align}
  \label{11}
  \beta^0 & =
    \begin{pmatrix}
      0 & 1 & 0 & 0 & 0\\
      1 & 0 & 0 & 0 & 0\\
      0 & 0 & 0 & 0 & 0\\
      0 & 0 & 0 & 0 & 0\\
      0 & 0 & 0 & 0 & 0
    \end{pmatrix},
  &
  \beta^1 & = 
  \begin{pmatrix}
    0 & 0 &-1 & 0 & 0\\
    0 & 0 & 0 & 0 & 0 \\
   -1 & 0 & 0 & 0 & 0\\
    0 & 0 & 0 & 0 & 0\\
    0 & 0 & 0 & 0 & 0
  \end{pmatrix},\\
  \beta^2 & =  
  \begin{pmatrix}
    0 & 0 & 0 &-1 & 0\\
    0 & 0 & 0 & 0 & 0\\
    0 & 0 & 0 & 0 & 0\\
   -1 & 0 & 0 & 0 & 0\\
    0 & 0 & 0 & 0 & 0 
  \end{pmatrix},
  &
  \beta^3 & =
  \begin{pmatrix}
    0 & 0 & 0 & 0 &-1\\
    0 & 0 & 0 & 0 & 0\\
    0 & 0 & 0 & 0 & 0\\
    0 & 0 & 0 & 0 & 0\\
   -1 & 0 & 0 & 0 & 0
  \end{pmatrix}.
\end{align}

The matrices ${\beta ^\mu }\left( x \right)$  in \eqref{6} are given more explicitly as
\begin{align}
  \label{12}
  \beta^t & =
  \begin{pmatrix}
    0 & 1 & 0 & 0 & 0\\
    1 & 0 & 0 & 0 & 0\\
    0 & 0 & 0 & 0 & 0\\
    0 & 0 & 0 & 0 & 0\\
    0 & 0 & 0 & 0 & 0
  \end{pmatrix},
                    &
  \beta^r & = 
  \begin{pmatrix}
    0 & 0 &-\cos\varphi & -\sin\varphi & 0\\
    0 & 0 & 0 & 0 & 0 \\
    -\cos \varphi & 0 & 0 & 0 & 0\\
    -\sin \varphi & 0 & 0 & 0 & 0\\
    0 & 0 & 0 & 0 & 0
  \end{pmatrix},\\
  \beta ^\varphi & = 
  \begin{pmatrix}
    0 & 0 & \frac{\sin \varphi }{\alpha  r} &-\frac{\cos\varphi }{\alpha r} & 0\\
    0 & 0 & 0 & 0 & 0\\
    \frac{\sin \varphi }{\alpha r} & 0 & 0 & 0 & 0 \\
    -\frac{\cos \varphi}{\alpha r} & 0 & 0 & 0 & 0\\
    0 & 0 & 0 & 0 & 0
  \end{pmatrix},
                  &
    \beta ^z & =
   \begin{pmatrix}
    0 & 0 & 0 & 0 &-1\\
    0 & 0 & 0 & 0 & 0\\
    0 & 0 & 0 & 0 & 0\\
    0 & 0 & 0 & 0 & 0\\
   -1 & 0 & 0 & 0 & 0
   \end{pmatrix}
\end{align}
where ${\beta^\circ },{\beta^r},{\beta^\varphi }$ and ${\beta^z}$
are the general form of the Kemmer matrices in this spacetime.

\section{ Solution of DKP oscillator in cosmic string background}

In this section, we concentrate our efforts in the interaction
called DKP oscillator.
For this external interaction we use the non-minimal substitution 
\begin{align}\label{13}
{\partial_r} \to {\partial_r} + M\omega r\eta^{0},
\end{align}
where $\omega$ is the oscillator frequency and
${\eta ^0} = 2{\beta ^0}^2 - 1$.
Considering only the radial component, with the non-minimal substitution
one gets
\begin{align}
  \label{14}
  \left[i{\beta ^ \circ }{\partial_t} + i{\beta ^r}({\partial_r}
  + M\omega r\eta^{0}) + i{\beta ^\varphi }({\partial_\varphi }
  - {\Gamma_\varphi }) + i{\beta ^z}{\partial_z}
  - M\right]\Psi (t,r,\varphi,z) = 0.
\end{align}
As the interaction is time-independent one can write the
spinor as
\begin{equation*}
  \Psi (t,r,\varphi ,z) = {e^{ - i(Et - m\varphi  - kz)}}\Psi (r),
\end{equation*}
  where
$E$ is  the energy of the scalar boson, and 
the five-component spinor can be written
as $\Psi(r) = \left(\psi_1(r),\ldots,\psi_5(r) \right)^{T}$
and the DKP equation for scalar bosons becomes
(for compactness of next equations, we momentarily drop the $r$
dependence in the spinor components)
\begin{align*}
 \label{15}
   {} & - m\sin\varphi\psi_1 - \alpha r M \psi_3
           - i \alpha r \cos\varphi
           \left[M \omega r \psi_1 + \psi_1^\prime \right]=0 ,\\
  {} & m\cos\varphi \psi_1 - \alpha r M\psi_4
           -i \alpha r \sin\varphi
           \left[M \omega r \psi_1 + \psi_1^\prime \right]=0,\\
   {} & k_z \psi_1 -M \psi_5=0,\\
   {} & E \psi_1 - M\psi_2=0,\\
   {} & \alpha r
           \left[
           - M \psi_1 + E\psi_2 + k_z\psi_5
           + i \cos\varphi(1 -\alpha + \alpha M \omega r^2)\psi_3
           - i m \psi_4 - \alpha r \psi_3^\prime 
           \right]\\
         & + i \sin\varphi
           \left[
           i m \psi_3
           + (1 - \alpha  + \alpha M \omega r^2)\psi_4
           - r\alpha \psi_4^\prime
           \right]=0.
\end{align*}
By solving the above system of equations in favour of $\psi_1(r)$ we get
\begin{align}
  \psi {}_2 = {}
  & \frac{{E{\psi_1}}}{{M}},\\
  {\psi_3} = {}
  & \frac{ - m\sin\varphi\psi_1
    + i [ - \alpha M \omega r^2 \cos\varphi\psi_1
    - \alpha r \cos\varphi\psi'_1]}
    {\alpha r M},\\
  {\psi_4} = {}
  & \frac{m\cos(\varphi ){\psi_1}
    + i [ - \alpha M \omega r^2 \sin\varphi\psi_1 -
    r\alpha \sin\varphi \psi'_1]}
    {\alpha r M},\\
    {\psi_5} = {}
  & \frac{{{k_z}{\psi_1}}}{{M}}.
\end{align}
Combining these results we obtain an equation of motion for the first
component of the DKP spinor
\begin{align}\label{17}
  \psi_1^{''}(r)
  + \frac{{\alpha-1}}{{\alpha r}}\psi_1^{'} (r) -
  \left(
  {E^2} - {M^2} + k_z^2 - \frac{{(2\alpha-1)M\omega }}{\alpha }
  + \frac{m^2}{\alpha^2r^2} + M^2\omega^2r^2
  \right)\psi_1(r) = 0.
\end{align}
In order to solve the above equation, we employ the change of variable:
$s = r^2$, thus we rewrite the radial equation \eqref{17} in the form 
\begin{equation}\label{18}
  \psi_1^{''}(s) + \frac{(2\alpha-1)}{2\alpha s}\psi_1^{'}(s)
  + \frac{1}{4s^2}
  \left(
     - \xi_1 s^2 +\xi_2 s- \xi_3
  \right)\psi_1(s) = 0.
\end{equation}
where
\begin{align}\label{19}
  \xi_1 = {} & M^2\omega^2,\\
  \xi_2 = {} & -E^2 - k_{z}^2 + M^2
               + \frac{(2\alpha-1)}{\alpha}M\omega, \\
  \xi_3 = {} & \frac{m^2}{\alpha^2}.
\end{align}
which gives the energy levels of the relativistic Klein-Gordon equation
from \cite{IJTP.48.337.2009}
\begin{multline}\label{20}
{\alpha_2}n - (2n + 1){\alpha_5} + (2n + 1)(\sqrt {{\alpha_9}}  + {\alpha_3}\sqrt {{\alpha_8}} ) + n(n - 1){\alpha_3} + {\alpha_7} + 2{\alpha_3}{\alpha_8} + 2\sqrt {{\alpha_8}{\alpha_9}}  = 0,
\end{multline}
where
\begin{align*}
  {\alpha_1} & = \alpha  - \frac{1}{2}, &
  {\alpha_2} & = 0, &
  {\alpha_3} & = 0, & 
  {\alpha_4} & = \frac{1}{2}\left(\frac{3}{2} - \alpha \right), &
  {\alpha_5} & = 0, \\
  {\alpha_6} & = {\xi_1}, & 
  {\alpha_7} & =  - {\xi_2}, &
  {\alpha_8} & = {\alpha_4}^2 + {\xi_3}, &
  {\alpha_9} & = {\xi_1}, &
  {\alpha_{10}} & = 1 + 2\sqrt {{\alpha_8}}, \\
  & &
  {\alpha_{11}} & = 2\sqrt {{\alpha_6}}, &
  {\alpha_{12}} & = {\alpha_4} + \sqrt {{\alpha_8}}, &
  {\alpha_{13}} & = - \sqrt {{\alpha_6}} .
\end{align*}
As the final step, it should be mentioned that the corresponding wave
function is
\begin{align}\label{22}
{\psi_1}(r) = {N r^{2{\alpha_{12}}}}{e^{{\alpha_{13}}{r^2}}}L_n^{{\alpha_{10}} - 1}({\alpha_{11}}{r^2}).
\end{align}
where $N$ is the normalization constant.


\section{ Solution of the DKP oscillator in presence linear interaction}

Let us now to analyse the situation when
a DKP field interacts with a scalar potential $U(r)$, which is
introduced via the substitution $M\rightarrow M+U(r)$.
Thus, \eqref{14} becomes
\begin{align}\label{23}
\left\{i{\beta ^ \circ }{\partial_t} + i{\beta ^r}({\partial_r} +
  M\omega r\eta^{0}) + i{\beta ^\varphi }({\partial_\varphi } -
  {\Gamma_\varphi }) + i{\beta ^z}{\partial_z} - [ M+U(r)]\right\}
  \Psi (t,r,\varphi,z) = 0.
\end{align}
Here we are interested in studying the linear scalar potential:
\begin{align}\label{24}
U(r) = a\,r.
\end{align}
In order to solve Eq. \eqref{23} we make the following change
of variables:
\begin{align}\label{25}
{\psi_1}(r) = {r^{\frac{{1 - \alpha }}{{2\alpha }}}}{(M + a\,r)^{\frac{1}{2}}}R(r).
\end{align}
This leads us to an equation without first-order derivative term
\begin{multline}\label{26}
  R^{''}(r) +
  \Bigg[ - {E^2} - {k_z}^2 + M \left(M + \omega  - \frac{\omega}{\alpha }\right)
  +  \frac{{{\alpha ^2} - 4{m^2} -1}}{{4{\alpha ^2}{r^2}}}
  - \frac{{a(1 - \alpha )}}{{2M\alpha r}} + 2aMr + \\
  ({a^2} - {M^2}{\omega ^2}){r^2}
  - \frac{{3{a^2}}}{{4{{(M + ar)}^2}}}
  + \frac{{a^2(1 - \alpha )
      + {2M^3\alpha}\omega }}{{2M\alpha(M + ar)}}\Bigg]R(r) = 0.
 \end{multline}
The next step, is to write  $R_{n,m  }(r )$ in the following form
\begin{align}\label{27}
{R_{n,m }}(r) = {R_n}(r ){e^{{g_m }(r )}}.
\end{align}
Thus, the $r$-dependent terms in Eq. \eqref{27} suggest that we take
${g_m }\left( r \right)$ as 
\begin{align}\label{28}
{g_m}(r) ={b_1}r + {b_2}{r^2} + {b_3} \log(r) + {b_4} \log(M + ar).
\end{align}
where the five constants $b_1, \ldots,b_4$ are to be expressed in terms
of the physical constants $\alpha$, $a$, $\omega$,  $M$, $M$ $m$, $k_z$
and $E$.
For nodeless states, with $n=0$, we have $R_n(r)=1$, consequently \cite{PRD.72.123003.2005,CQG.23.2653.2006,NPB.662.3.2003}
\begin{align}\label{29}
{R_{0,m}}(r) = {e^{{g_m }(r)}},
\end{align} 
In this manner, from Eq. \eqref{20} we have
\begin{align}\label{30}
R''_{0,m}(r) + ( - g_m'' - {g_m'^2}){R_{0,m }}(r) = 0.
\end{align}
Thus substituting Eq. \eqref{28} into Eq. \eqref{30}, the latter equation
becomes 
\begin{multline}\label{31}
  {{R''}_{0,m}}(r) +
  \Bigg( - b_1^2 - 2{b_2} - 4r{b_1}{b_2}
  - 4{b_2}{b_3} - 4{b_2}{b_4} - 4b_2^2{r^2}
  + \frac{{{b_3} - b_3^2}}{{{r^2}}} - \\
  \frac{2 M b_1 b_3 + 2 a b_3 b_4}{M r} + \frac{{{a^2}{b_4}
      - {a^2}b_4^2}}{{{{(M + ar)}^2}}}
  + \frac{4M^{2}b_2b_4 - 2 M a b_1 b_4
      + 2 a^2 b_3 b_4}{M(M + ar)}
  \Bigg){R_{0,m}}(r) = 0.
\end{multline} 
If we compare Eq. \eqref{31} with Eq. \eqref{26}, we acquire the
following six equations (displayed here along with their respective
factors in ($r$):
\begin{align}\label{32}
  {} & - {E^2} - {k_z}^2 + M(M + \omega  - \frac{\omega }{\alpha })
        + b_1^2 + 2{b_2} + 4r{b_1}{b_2} + 4{b_2}{b_4} + 4{b_2}{b_3}=0,\nonumber\\
 {} & - 2a{b_1}{b_4} + 4M{b_2}{b_4} + 2{a^2}{b_3}{b_4}
        + \frac{{{a^2}( - 1 + \alpha )}}{{2M\alpha }} - {M^2}\omega=0,\nonumber\\
 {} & \frac{{2a{b_3}{b_4}}}{M} + 2{b_1}{b_3}
        + \frac{{a( - 1 + \alpha )}}{{2M\alpha }}=0,\nonumber\\
  {} & \frac{{ - 1 + {\alpha ^2} - 4{m^2}}}{{4{\alpha ^2}}}
        - {b_3} + b_3^2=0,\nonumber\\
  {} & ({b_4} - b_4^2) + \frac{3}{4}=0,\nonumber\\
  {} & 4b_2^2 + ({a^2} - {M^2}{\omega ^2})=0, \nonumber\\
  {} & 4{b_1}{b_2} + 2aM = 0.
\end{align}

These equations can be solved for $b_1$, $b_2$, $b_3$ and $b_4$ they
also provide constraints on the physical parameters, in particular, the
energy $E$. 
Equation \eqref{32} admit the following solutions:
\begin{align}\label{33}
  {b_1} = {} & \mp \frac{{aM}}{{\sqrt { - {a^2} + {M^2}{\omega ^2}} }},\nonumber\\
  {b_2} = {} &  \pm \frac{1}{2}\sqrt { - {a^2} + {M^2}{\omega ^2}}, \nonumber\\
  {b_3} = {} &  \pm \frac{{{\alpha ^2} + \sqrt {{\alpha ^2}
               + 4{\alpha^2}{m^2}} }}{{2{\alpha ^2}}},\nonumber\\
  {b_4} = {} & \frac{1 \pm 2}{2}.
\end{align}
Finally, the solution for DKP oscillator interacting with a linear
scalar potential can be written as
\begin{align}\label{34}
\Psi _{n,m} (t,r,\varphi ,z)=N_{n,m} {e^{i({k_z} + m\varphi  - Et)}}{e^{{b_1}r + {b_2}{r^2}}}{r^{{b_3} + \frac{{1 - \alpha }}{{2\alpha }}}}{(M + ar)^{{b_4} + \frac{1}{2}}},
\end{align}
where $N_{n,m}$ is normalization constant.

\section{Conclusion}

In this contribution, we have investigated the relativistic quantum
dynamics of DKP oscillator in the presence of the linear interaction, on
the curved spacetime of a cosmic string. 
From the corresponding DKP equation, we analyze the influence of the
topological defect on the equation of motion, the energy spectrum and
the wave function.
We established and found solutions of covariant DKP oscillator in cosmic
string background.
We present solution of DKP oscillator presence  linear interaction.
We solved the DKP oscillator analytically by using a proper ansatz solution
to the equation. 

\bibliographystyle{unsrt}

\end{document}